\documentclass[sigconf]{acmart}

\AtBeginDocument{%
  }

\setcopyright{rightsretained}
\copyrightyear{2023}
\acmYear{2023}
\acmDOI{}
\acmBooktitle{KDD Cup 2023 Workshop: Multilingual Session Recommendation Challenge}

\acmConference[KDDCup '23]{}{August 09, 2023}{Long Beach, USA}
\acmPrice{}
\acmISBN{}

\begin{document}

\title{A Completely Locale-independent Session-based Recommender System by Leveraging Trained Model}

\author{Yu Tokutake}
\email{tokutakeyuu@uec.ac.jp}
\affiliation{%
  \institution{The University of Electro-Communications}
  \city{Tokyo}
  \country{Japan}
}

\author{Chihiro Yamasaki}
\email{yamasaki@uec.ac.jp}
\affiliation{%
  \institution{The University of Electro-Communications}
  \city{Tokyo}
  \country{Japan}
}

\author{Yongzhi Jin}
\email{jin-yongzhi@uec.ac.jp}
\affiliation{%
  \institution{The University of Electro-Communications}
  \city{Tokyo}
  \country{Japan}
}

\author{Ayuka Inoue}
\email{ayuka.i@uec.ac.jp}
\affiliation{%
  \institution{The University of Electro-Communications}
  \city{Tokyo}
  \country{Japan}
}

\author{Kei Harada}
\authornote{Corresponding author.}
\email{harada@uec.ac.jp}
\affiliation{%
  \institution{The University of Electro-Communications}
  \city{Tokyo}
  \country{Japan}
}

\renewcommand{\shortauthors}{Y. Tokutake, C. Yamasaki, Y. Jin, A. Inoue, and K. Harada}

\begin{abstract}
  In this paper, we propose a solution that won the 10th prize in the KDD Cup 2023 Challenge Task 2 (Next Product Recommendation for Underrepresented Languages/Locales). 
Our approach involves two steps:
(i) Identify candidate item sets based on co-visitation, and
(ii) Re-ranking the items using LightGBM with locale-independent features, including session-based features and product similarity.
The experiment demonstrated that the locale-independent model performed consistently well across different test locales, and performed even better when incorporating data from other locales into the training.
\end{abstract}




\maketitle

\section{Introduction} \label{sec:introduction}
The Amazon KDD Cup 2023\footnote{\url{https://www.aicrowd.com/challenges/amazon-kdd-cup-23-multilingual-recommendation-challenge}} presents challenges that are e-commerce shopping session recommendations on the Amazon-M2 dataset \cite{jin2023amazon}.
This competition focuses on session-based recommender system (SBRS) in multilingual and imbalanced scenarios and consists of three tasks.
Participants predicted the next item (Tasks 1 and 2) or the title of the next item (Task 3) to be purchased based on the previous session item set.
Our focus is on the next product prediction solution.

SBRS predicts the next item based on the user interaction sequence.
In SBRS, sequential dependencies among the items in a session are highly relevant to recommendation contents \cite{2021jul_s.Wang}. Therefore, SBRS aims to model users' short-term and dynamic preferences, unlike collaborative and content-based filtering, which capture long-term preferences \cite{2021jul_s.Wang, 2018dec_m.Ludewig}.
E-commerce is one of the most popular cases of SBRS, which identifies a user's explored topic and provides relevant items from numerous number items published on a site \cite{2018dec_m.Ludewig}.
SBRS predicts a user's preference for the next item based on a series of user activities, such as clicking, viewing, and purchasing items on a website \cite{2021jul_s.Wang, 2017nov_c.Wu}.
The relationship between these interactions is useful for estimating a user's search intentions and interesting topics \cite{2017nov_c.Wu}.
Traditional approaches used in SBRS are based on Markov chains and k-nearest neighbor.
Markov chains model the transition relationship between interactions. Rendle et al. and He et al. combined matrix factorization and Markov chains \cite{2010apr_s.Rendle, 2016sep_r.he}.
$k$-nearest neighbor uses the similarity between interactions. Grag et al. proposed STAN \cite{2019jul_d.Garg}, which includes both sequential and time information.
Recent methods are mostly based on recurrent neural network (RNN) because of its advantages in modeling sequential relationships \cite{2016sep_b.Twardowski, 2017aug_d.Jannach}.
GRU4Rec \cite{2015nov_b.Hidasi} is a representative RNN-based method that introduced Gated Recurrent Units. 
Quadrana et al. enhanced GRU4Rec by incorporating hierarchical relationships across sessions \cite{2017aug_m.Quadrana}.
In addition, assuming an anonymous user who does not log in, some studies have proposed methods that focus on a single session and are not linked to user information \cite{2017nov_c.Wu, 2015nov_b.Hidasi, 2016sep_b.Twardowski}.
It is important for such a setting to capture the changes in user preferences and interaction transitions in the shorter term.

In this paper, we propose an approach for predicting the next product in a single session. We used Item2Vec similarity \cite{2016may_o.Barkan} and Universal Sentence Encoder similarity \cite{2018mar_d.Cer, 2019jul_y.Yang, 2019aug_m.Chidambaram} in addition to the co-visitation relations between products, to better capture the transition relationships between products.
To handle massive numbers of candidate products, we first extracted several hundred candidate products and then re-ranked them using LightGBM.

The remainder of this paper is organized as follows.
Section \ref{sec:task_description} provides information on the Amazon-M2 dataset and evaluates the recommended results.
In Section \ref{sec:method}, we introduce the proposed solution.
Section \ref{sec:experiment} presents our experimental results.
\section{Task Description} \label{sec:task_description}
\subsection{Dataset}
Amazon-M2 dataset consists of two types: user sessions and product attributes.
These are multilingual data from six different locales: English (UK), German (DE), Japanese (JP), French (FR), Italian (IT), and Spanish (ES).
The statistics of the dataset are presented in Table \ref{table:dataset_statistics}.
In this paper, we will refer to three locales with sufficient data (DE, JP, and UK) as ``SL'' and to other locales with insufficient data (ES, FR, and IT) as ``IL.''

User sessions are divided into train and test sessions.
Sessions are made up of locale and chronologically ordered item sets (prev\_items) and are not personally identifiable.
The length of prev\_items is two or more and different from each session.
In the case of train sessions, it also has one item next to prev\_items (next\_item).
The test sessions contain no information regarding the next\_item.
Task 1 was the next product prediction for the test sessions of the SL, and Task 2 was for that of the IL.
For both tasks, 100 candidate items of the next\_item were listed for each session.

Product attributes include records such as the title, price, and brand.
Participants can use them for recommendations.

\begin{table}[t]
  \centering
  \caption{Statistics of Amazon-M2 dataset}
  \label{table:dataset_statistics}
  \begin{tabular}{ccc}
    \hline
    Locale & \# Sessions & \# Products \\
    \hline
    DE & 1,111,416 & 513,811 \\
    JP & 979,119 & 389,888 \\
    UK & 1,182,181 & 494,409 \\
    ES & 89,047 & 41,341 \\
    FR & 117,561 & 43,033 \\
    IT & 126,925 & 48,788 \\
    \hline
  \end{tabular}
\end{table}

\subsection{Evaluation}
The next product prediction was evaluated using the mean reciprocal rank (MRR), which is used in many SBRS studies \cite{2016sep_b.Twardowski, 2017aug_m.Quadrana, 2017aug_d.Jannach, 2018dec_m.Ludewig, 2019jul_d.Garg, 2019nov_r.Qiu}.
MRR is calculated as
\begin{align*}
    \mathrm{MRR}@K = \frac{1}{N}\sum_{t \in T}\frac{1}{\mathrm{Rank}(t)}
\end{align*}
where $T$ is the test session set, $N$ is the length of the set ($N = |T|$), and $\mathrm{Rank}(t)$ is the rank where the ground truth (next\_item) appears in the top-$K$ recommended list for the test session $t$ (if the top-$K$ list does not have the next\_item, $1 / \mathrm{Rank}(t) = 0$).
For these tasks, $K$ was set to 100.

\section{Method} \label{sec:method}

\subsection{Basic Strategy}
The tabular data approach was chosen for this task, that is, 
we extracted features from sessions and items and built a supervised model to predict whether an item was the next\_item or not.
The reason for this choice was that the matrix factorization approach, which is commonly used in recommendation tasks, does not work well.
We believe this is because the data were too sparse to be represented in 100 or 300 dimensions.

The candidate items were selected using the co-visitation features for each session before training the model.   
Ideally, we should predict for every session and every item. However, this is too much to do not only for task 1 with large data, but also for task 2 with relatively small data. 

LightGBM was used for binary classification, and the output probability for ranking the items was used. 
If there were fewer than 100 candidate items, they were added using the Item2Vec and Universal Sentence Encoder model.

\begin{figure*}[t]
    \centerline{\includegraphics[width=15.0cm]{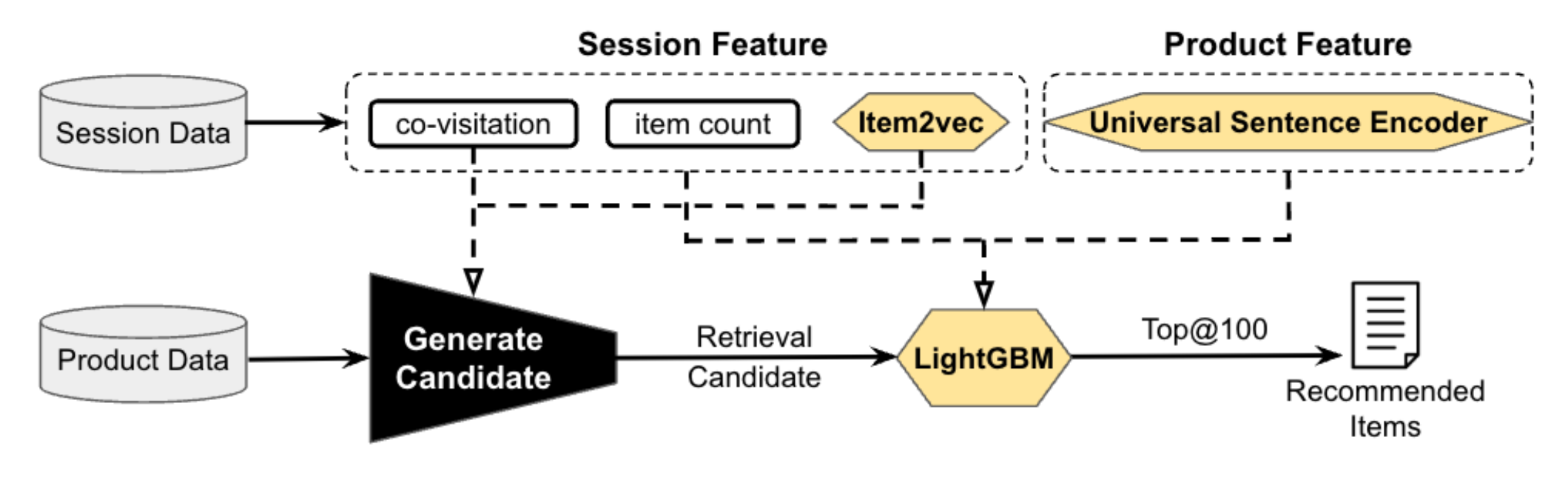}}
    \caption{The overview of our approach}
    \label{figure:overview}
\end{figure*}

An overview of the proposed approach is presented in Figure \ref{figure:overview}.

\subsection{Feature Extractions}

We have four groups of features.

\subsubsection{Co-Visitation based features} 

We created multiple versions of concurrent visits in prev\_items; This is not simply the number of co-occurrences of items A and B (we denote by $c_s(A, B)\,$), but the number of B after A ($c_{axb}(A, B)\,$), B immediately after A ($c_{ab}(A, B)\,$), and B immediately before A ($c_{ba}(A, B)\,$).
Of course, we also use co-visitation information, including the next\_item. 
The interaction of item A in prev\_items and the next\_item B are counted in two ways: 
The count of occurrences where A is in previous items (we denote by $y_{axb}(A, B)\,$) and B is the next\_item and that A is the last item in previous items and B is the next\_item ($y_{ab}(A, B)\,$). 

For example, if a session has A, B, A, C as prev\_items and D as next\_item, $c_s(A, B)$ is added by 2, $c_{axb}(A, B)$ is added by 1, $y_{axb}(A, D)$ is added by 2, and $y_{axb}(C, D)$ is added by 1.

%
Note that the actual calculation was performed by the per-session cross-join, not on every item pair. 
Thus, the calculations are feasible.  
Note that the same items but different locales were treated as different items.

These features are expected to be the numerator of similarity, and the similarity between all items in prev\_items and a candidate item can be obtained with a simple data join operation. 
We narrowed down the calculations to the last few items in the session to limit the number of columns in the table.
For example, limiting it to the last 10 items means that we obtained 60 similarity features. 

\subsubsection{Item count features} 
These features are expected to be the denominators for co-visitation-based similarity discussed above. 
We simply counted how many times it appeared in prev\_items and how many times it appeared as next\_item. 

Since we took a tabular data approach, for each session and a candidate item, count features for the candidate item and last few items extracted from prev\_items.

\subsubsection{Text based similarity}
A quick look at the data reveals that there are nearly identical products, such as different sizes, different colors, and different quantities, and that there are many sessions where the next\_item is almost the same as one of the prev\_items. 
We must define the similarity between items based on product information to take advantage of this property.

We used the Universal Sentence Encoder (USE) \footnote[2]{https://tfhub.dev/google/universal-sentence-encoder-multilingual-large/3} for this; 
This is because we wanted to use language-agnostic features as much as possible and because we felt the data were not big enough to learn language embedding.

Similarity calculation was performed in a simple manner. 
Combining product information strings with the conjunction ":" and embedding them with USE yields a 512-dimensional vector, and the similarity between the two items is defined by the Euclidean inner product.

\subsubsection{SBRS based prediction}
We used the Item2Vec model used in session-based recommendations to capture the relationships between items across sessions. The models were trained per locale.

They were implemented using gensim.
The hyper-parameters were determined for each locale and tuned for the number of vector dimensions [25, 50, 75, 100] and the subsampling thresholds [0.01, 0.001, 0.0001].
These values are listed in Table \ref{table:item2vec_param}.
Gensim default values were used for all other parameters.

\subsection{Cross Validation}
For parameter tuning, we used 5-fold cross-validation (CV) based on the sequential number of sessions. 
Simple cross-validation alone is not enough because some features are created by looking at the next\_item.
$y_{axb}$, $y_{ab}$, the count features for the next\_item, and Item2Vec-based predictions have the potential for severe overfitting, similar to target encoding. 

To address this issue, these features were calculated outside the fold to which the session belonged; that is, we had five versions for each feature, and for each record, the version that did not use the record was used.

\subsection{Candidates Generation}
Co-Visitation-based features were used in the candidate generation process.
All items suggested by one of the co-visitation-based features were considered candidates.
More precisely, item B is a candidate for the session if $c_s(A, B) > 0$ or $y_{ab}(A, B) > 0$ for some item A in the session. 

After candidate items were picked, text-based similarity was computed between the candidate items and the last 10 items of prev\_items for each session.
That is, we obtained 10 similarity features for each candidate item.

Item2Vec-based predictions also suggested items; however, we decided to simply use them as features because the co-visitation-based features scored much higher. 
1) If an item is a candidate in both methods, the prediction is used as a feature.
2) If an item is a candidate in a co-visitation-based manner and not in Item2Vec-based predictions, null is set. 
3) If an item is a candidate in Item2Vec-based predictions and not in a co-visitation-based manner, the item is not a candidate in the first place. 
As an exception, if candidate items are less than 100, Item2Vec-based predictions are used to fill up to 100. 
If the number of candidate items is still less than 100 after filling in the Item2Vec-based predictions, USE-based predictions, which recommend items with high text similarity to the last item in prev\_items, are used to fill up to 100.

\begin{table}[t]
  \centering
  \caption{Item2Vec parameters used for each locale}
  \label{table:item2vec_param}
  \begin{tabular}{ccc}
    \hline
    Locale & Vector Dimensions & Subsampling Thresholds \\
    \hline
    DE & 100 & 0.001 \\
    JP & 100 & 0.0001 \\
    UK & 100 & 0.0001 \\
    IT & 100 & 0.001 \\
    FR & 75 & 0.001 \\
    ES & 100 & 0.0001 \\
    \hline
  \end{tabular}
\end{table}

\section{Experiment} \label{sec:experiment}
Our primary goal was to improve the score of Task 2, and we applied the pre-trained LightGBM model obtained in the process to Task 1.
The results of the leaderboard MRR comparison for various training data locales for each task are listed in Table \ref{table:mrr_comparison}.
(IL + DE) + (IL + JP) + (IL + UK) is an ensemble of IL + DE, IL + JP, and IL + UK.
In Table \ref{table:mrr_comparison}, the best values are in bold.
In Task 2, we can see that using data from other SL as training data in addition to IL improved the performance by up to 0.88\%.
The results for IL with only one of the SL and their ensembles showed no significant change.
In contrast, in Task 1, the results for SL-only were not good, and the IL-only model for Task 2 was better.
Applying the same model with local-independent features to both tasks got some results, especially in Task 2, where we won the 10th place.
Furthermore, while there may be some influence of experimental settings and sampling of candidate items, the result of SL-only and IL-only in Task 1 suggests that large training data may not be necessary and that small data transfers may be enough.

\begin{figure}[t]
  \centerline{\includegraphics[height=7.2cm]{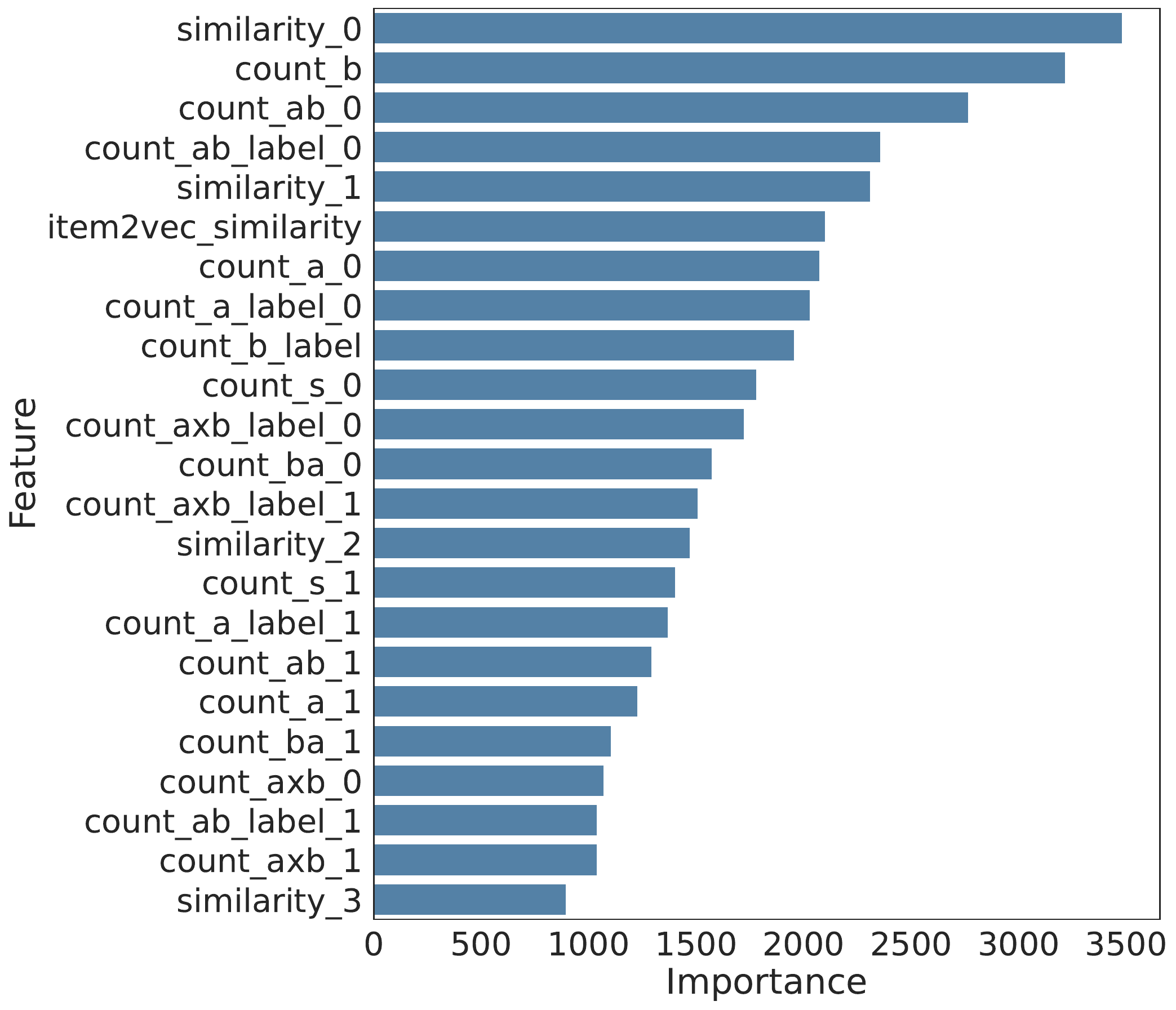}}
  \caption{The top-20 feature importance of LightGBM model}
  \label{figure:feature_importance}
\end{figure}

The top-20 feature importance of the LightGBM model is presented in Figure \ref{figure:feature_importance}.
From Figure \ref{figure:feature_importance}, it is clear that features of the new item in prev\_items and the similarity features have high importance.
In addition, item2vec\_similarity, which is an SBRS-based feature, exhibited a high value.

\begin{table}[t]
  \centering
  \caption{Leaderboard MRR@100 comparison for various training data locales for each task}
  \label{table:mrr_comparison}
  \begin{tabular}{ccc}
    \hline
    Training Data Locales & Task 1 & Task 2 \\
    \hline
    IL & 0.37584 & 0.43718 \\
    SL & 0.37113 & 0.41827 \\
    IL + DE & 0.37893 & 0.44008 \\
    IL + JP & 0.38010 & 0.44007 \\
    IL + UK & \textbf{0.38055} & 0.43996 \\
    (IL + DE) + (IL + JP) + (IL + UK) & 0.36975 & 0.44007 \\
    IL + SL & 0.37065 & \textbf{0.44101} \\
    \hline
  \end{tabular}
\end{table}

\section{Conclusion} \label{sec:conclusion}
In this study, we present our solution for the Amazon KDD Cup 2023.
We propose a completely locale-independent model, which generates a candidate item for a session and predicts whether the item is the next\_item or not, based on four types of features.
The experimental results demonstrated that incorporating data from locales with ample training data into data from locales with limited training data enhanced recommendation accuracy.

In future studies, we aim to implement two improvements to our solution model.
Firstly, given the significant number of candidate items, particularly in Task 1, we plan to explore more efficient methods for narrowing down the selection.
This investigation will lead to further enhancement in the system's performance and scalability.
Secondly, we intend to enrich the model by incorporating features from other state-of-the-art session-based recommendation systems (SBRS) like GRU4Rec, as introduced in Section \ref{sec:introduction}.
By integrating these features and constructing ensembles based on diverse SBRS methods, we expect to achieve improved recommendation accuracy and enhanced robustness in our predictions.

\begin{acks}
We would like to thank Dr. Kazushi Okamoto for his help in writing this paper.
This work is supported by Design Thinking and Data Science Program at The University of Electro-Communications.
\end{acks}

\bibliographystyle{ACM-Reference-Format}
\bibliography{references}


\end{document}